\documentclass[amsmath,amssymb,showkeys,superscriptaddress,floatfix,prd,10pt,aps,twocolumn]{revtex4-2}
\usepackage{graphicx,epstopdf}
\usepackage[caption=false]{subfig}
\usepackage{multirow}
\usepackage{appendix}
\def\xslash{x\!\!\!\slash }
\def\kslash{k\!\!\!\slash }

\usepackage{colortbl}
\usepackage{xcolor}
\usepackage[colorlinks=true, urlcolor=blue, linkcolor=green, citecolor=green]{hyperref}

\begin{document}

\title{Molecular Interpretation of $X(3960)$ as $D_s^+ D_s^-$ State} % Force line breaks with \\

\author{Halil Mutuk}
 \email{hmutuk@omu.edu.tr}
 \affiliation{Department of Physics, Faculty of Sciences, Ondokuz Mayis University, 55139 Samsun, T\"{u}rkiye}
 \affiliation{Institute for Advanced Simulation, Institut für Kernphysik and Jülich Center for Hadron Physics, Forschungszentrum Jülich, D-52425 Jülich, Germany}

%\date{\today}
 
\begin{abstract}
We study  $D_s^+ D_s^-$ and $D \bar D$ states assuming that they are hadronic molecules with $J^{PC}=0^{++}$ quantum number. We use two-point QCD sum rule formalism and extract the mass and decay constant values of these states. We take into account contributions of various quark, gluon, and mixed vacuum condensates up to dimension eight. The extracted mass and decay constant values of $D \bar D$ and $D_s^+ D_s^-$ states read as $M_{D \bar D} = 3795^{+85}_{-82} ~\mathrm{MeV} $, $f_{D \bar D} = 1.70^{+0.33}_{-0.29} \times 10^{-2}  ~\mathrm{GeV}^5$, and $M_{D_s^+ D_s^-} = 3983^{+93}_{-88} ~\mathrm{MeV}$, $f_{D_s^+ D_s^-} = 2.52^{+0.64}_{-0.54} \times 10^{-2}  ~\mathrm{GeV}^5$, respectively. The predicted mass of $D_s^+ D_s^-$ state is in good agreement with the recent LHCb observation and supports quantum number and molecular picture assignments. A possible observation of $D \bar D$ state would help for establishing the lowest four-quark state in charmonium sector.
\end{abstract}

%\keywords{Magnetic and quadrupole moments, electromagnetic form factors,  open-flavor tetraquark  states,   QCD light-cone sum rules}

\maketitle

\section{\label{sec:level1}Prologue}
Multiquark states have been proposed when the quark model was introduced by Gell-Mann \cite{Gell-Mann:1964ewy} and Zweig \cite{Zweig:1964ruk} more than fifty years ago. After the proposition of quark model, experimentally observed states fall in the scope of baryons $(qqq)$ and mesons $(q \bar q)$. Theoretical investigations which are supported by the experimental observations of these traditional hadrons were the success of the quark model. This situation turned into a different scene when the observation of $\chi_{c1}(3872)$ (formerly known as $X(3872)$) state announced by the Belle Collaboration in 2003 \cite{Belle:2003nnu}. Apart from its mass which is lower than the quark model prediction, this state decays into $J/\psi \rho$ and $J/\psi \omega$ which are forbidden and OZI-suppressed for a bare $c\bar c$ configuration. Many more states are observed such as $Z_c(3900)$ \cite{Belle:2011aa}, $Z_c(4020)$ \cite{BESIII:2013ouc,BESIII:2013mhi}, $P_c$ \cite{LHCb:2015yax,LHCb:2019kea} and $Z_{cs}(3985)$ \cite{BESIII:2020qkh} whose physical properties cannot be obtained within the quark model framework. 

%The definition of exotic states can be done in terms of quantum numbers, electric charge and quark flavor. For example there is no coupling between spin and orbital angular momentum of a traditional meson which gives $J^{PC}=0^{--}, 2^{+-}, 2^{-+}$. If a meson has one of these quantum numbers, it must have an exotic nature. Furthermore, observed decay channels may also indicate the exotic nature for quark content. 

Experimentally known charmonium spectrum lies in the range of 3.0-4.5 GeV. The spectrum around 3.9 GeV would correspond to $\chi_{cJ}(2P)$ multiplet since the spectrum of $\chi_{cJ}(1P)$ is in the range of 3.4-3.5 GeV and is quite interesting because most of the experimentally observed charmoniumlike states are near the $\bar D^{*0}D^0/\bar D^{0}D^{*0}$ threshold for $\chi_{c1}(3872)$, $\bar D^{*}D/\bar DD^{*}$ threshold for $Z_c(3900)$, $\bar D^{*}D^{*}$ threshold for  $Z_c(4020)$ state, and $\bar D_s D^*/\bar D_s^* D$ threshold for  $Z_{cs}(3985)$ state. 

In this respect, experimental observations put some puzzles through our understanding of exotic states. In 2004 summer, the Belle Collaboration observed a state with a mass around 3940 MeV in the invariant $\omega J/\psi$ mass distribution of the $B \to K \omega J/\psi$ \cite{Belle:2004lle}. The BaBar Collaboration confirmed this state in the $B^{0,+} \to J/\psi \omega K^{0,+} $ but with a different mass around 3915 MeV \cite{BaBar:2007vxr}. Determination of the quantum numbers for this state came a couple of years later after the observation as $J^{PC}=0^{++}$ by BaBar Collaboration \cite{BaBar:2012nxg}. In 2020, the LHCb Collaboration found a structure named $\chi_{c0}(3915)$, in the $D^+D^-$ mass distribution of $B^+ \to D^+ D^- K^+$ decay with the mass and width of \cite{LHCb:2020bls,LHCb:2020pxc}

\begin{equation}
J^{PC}=0^{++}; ~ M=3924 \pm 2  ~ \text{MeV}; ~ \Gamma=17 \pm 5 ~ \text{MeV}. 
\end{equation}
Particle Data Group refer these states as $\chi_{c0}(3915)$ \cite{ParticleDataGroup:2020ssz}. This state is attributed as conventional $\chi_{c0}(2P)$ charmonium state in Refs. \cite{BaBar:2012nxg,Duan:2020tsx,Duan:2021bna}. Ref. \cite{Li:2015iga} argued $\chi_{c0}(3915)$ as an $S$ wave $D_s \bar D_s$ molecular state with a binding energy of 18 MeV. It was mentioned in Ref. \cite{Gonzalez:2016fsr} that experimental significance of the OZI forbidden $X(3915) \to \omega J/\psi $ decay could put a question mark to the conventional charmonia description of this state.

Very recently, the LHCb Collaboration \cite{results} announced the observation of a new structure named as $X(3960)$ in the $D_s^+ D_s^-$ mass distribution of the $B^+ \to D_s^+ D_s^- K^+$ decay with the mass and width of
\begin{equation}
J^{PC}=0^{++}; ~ M=3955 \pm 6 \pm 11 ~ \text{MeV}; ~ \Gamma=48 \pm 17 \pm 11 ~ \text{MeV}. 
\end{equation}
The assigned quantum number makes this structure an $S$ wave state. The resonant peak is just above the $D_s^+ D_s^-$ threshold which is 3937 MeV and is a good candidate for a $D_s^+ D_s^-$ molecule. The threshold of $D \bar D$ is 3739 MeV. According to quark model calculations \cite{Barnes:2005pb,Li:2009zu}, $\chi_{c0}(2P)$ states have masses around 3.9 GeV and $\chi_{c0}(3P)$ states have masses around 4.2 GeV. The quantum numbers, masses and widths of $\chi_{c0}(3915)$ and $X(3960)$ suggest these particles as the same particle. However, the branching ratio of these states
\begin{equation}
\frac{\Gamma (X \to D^+ D^-)}{\Gamma (X \to D_s^+ D_s^-)}=0.29 \pm 0.09 \pm 0.10 \pm 0.08
\end{equation}
says that $\Gamma (X \to D^+ D^-) < \Gamma (X \to D_s^+ D_s^-)$. This implies the exotic nature of $X(3960)$ state because conventional charmonium states predominantly decay into $D \bar D$ and $D^* \bar{D}^*$ states. Furthermore it is harder to excite an $s \bar s$ pair from vacuum compared with $u \bar u$ or $d \bar d$ pairs. Therefore, it is natural to expect that $\Gamma (X \to D^+ D^-) $ should be bigger than $\Gamma (X \to D_s^+ D_s^-)$. 

The $D \bar D$ and $D_s^+ D_s^-$ molecular states have been studied in various approaches before \cite{Gamermann:2006nm,Nieves:2012tt,Hidalgo-Duque:2012rqv,Prelovsek:2020eiw,Meng:2020cbk,Dong:2021juy} and after \cite{Bayar:2022dqa,Ji:2022uie,Xin:2022bzt,Xie:2022lyw,Guo:2022crh,Guo:2022zbc} the observation of $X(3960)$. In Refs. \cite{Gamermann:2006nm,Nieves:2012tt,Hidalgo-Duque:2012rqv}, a $D \bar D$ bound state was found whereas no bound state was found close to the $D_s^+ D_s^-$ threshold. If the $\chi_{c0}(3915)$ state couple both the $D^+ D^-$ and $D_s^+ D_s^-$, it can produce an enhancement close to the $D_s^+ D_s^-$ threshold and would appear as a signal in the $D_s^+ D_s^-$ mass distribution. In that case, the experimental observation of  $X(3960)$ would be explained without introducing an extra resonance. Refs. \cite{Prelovsek:2020eiw,Bayar:2022dqa} support this conclusion. In Ref. \cite{Ji:2022uie}, using nonrelativistic effective field theory it was found that a bound state or a virtual state below the $D_s^+ D_s^-$ mass threshold is needed to describe the $D_s^+ D_s^-$ mass distribution. Ref. \cite{Xin:2022bzt} studied $X(3960)$ as $D_s^+ D_s^-$ molecule  with $J^{PC}=0^{++}$ quantum number in QCD sum rule framework. Production rates of $D \bar D$ and $D_s^+ D_s^-$ states were studied via effective Lagrangian approach \cite{Xie:2022lyw}. Using improved chromomagnetic interaction model, $S-$wave $Qq \bar{Q} \bar{q}$ tetraquark states were studied in Ref. \cite{Guo:2022crh}. They observed that $X(3960)$ state can be well explained in their model.  In Ref. \cite{Guo:2022zbc} it was found that $\chi_{c0}(2P)$ state below the $D_s\bar{D}_s$ threshold can depict the structure of $X(3960)$. In one-boson-exchange model $D_s \bar{D}_s / D^{*} \bar{D}^{*} / D_s^* \bar{D}_s^*$ states were studied considering the $S-D$ wave mixing effects \cite{Chen:2022dad}. 

In this present study, we calculate the mass and decay constant of the $D^+ D^-$ and $D_s^+ D_s^-$ states in the context of the two-point QCD sum rule (QCDSR) \cite{Shifman:1978bx,Shifman:1978by}. We choose scalar interpolating current in correlation function to account the quantum numbers of the $\chi_{c0}(3915)$ and $X(3960)$ state. We evaluate the two-point correlation function in terms of various quark, gluon and mixed vacuum condensates. 

The paper is organized in the following way: in Section \ref{sec:level2} we provide the method of QCDSR calculations for the considered states to extract mass and coupling constant values. Section \ref{sec:level3} presents the details of numerical analyses and results for spectroscopic parameters. Section \ref{sec:level4} is reserved for summary and conclusion.  

\section{\label{sec:level2}Two-Point Correlator}
In order to extract mass and decay constant from QCDSR technique, an interpolating current representing the physical state in the problem should be constructed. To this end, we use the following interpolating currents interpreting $D \bar D$ and $D_s^+ D_s^-$ states as scalar molecules with quantum number $J^{P}=0^{++}$
\begin{eqnarray}
j_1(x)&=&\bar q^a(x) \gamma_{\mu} c^a(x) \bar c^b \gamma^{\mu} q(x)^b,  \\
j_2(x)&=&\bar s^a(x) \gamma_{\mu} c^a(x) \bar c^b \gamma^{\mu} s(x)^b, 
\end{eqnarray}
where $q=u, d$, The sum rules for masses and decay constants of the $D \bar D$ and $D_s^+ D_s^-$  states related to these currents are constructed from the two-point correlation function
\begin{equation}
\Pi(q) =  i \int d^4x e^{iqx} \langle 0 \vert T [j(x) j^\dagger (0)] \vert 0 \rangle, \label{corfunc}
\end{equation}
where $j(x)$ is the interpolating current, $T$ is the time ordered product between the operators and $\langle \cdots \rangle$ is the QCD vacuum expectation value. 

The basic idea of the QCDSR framework is that there is an interval in momentum $q$ where the correlation function may be equivalently described at both quark-gluon (QCD or OPE) and hadron (phenomenological) levels. A typical QCDSR calculation has two steps. In the first step, correlation function in the phenomenological side is represented by the hadronic degrees of freedom. Inserting complete set of intermediate states for the hadron of interest into the correlation function Eq. (\ref{corfunc}) and isolating ground state contribution of higher states and continuum, one can obtain
\begin{equation}
\Pi^{\text{Phen}}(q)=\frac{\langle 0 \vert j \vert H(p)\rangle \langle H(p) \vert j^\dagger \vert 0 \rangle}{m_H^2-q^2} + \cdots . \label{corphen0}
\end{equation}
Let $m_H$ and $f_H$ be the mass and decay constant of the hadron, respectively. One can write the matrix element as
\begin{equation}
\langle 0 \vert j \vert H \rangle = m_H f_H.
\end{equation}
With this definition  $\Pi^{\text{Phen}}(q)$ as can be written as
\begin{equation}
\Pi^{\text{Phen}}(q)=\frac{m_H^2 f_H^2}{m_H^2-q^2}+ \cdots, \label{corphen1}
\end{equation}
where $\cdots$ represents contributions of higher states. As a result of writing the spectral function in terms of intermediate states for the hadron of interest, all these states that couple to the interpolating current $j(x)$ contribute to the correlation function. In the case of four-quark mesons, it should be mentioned that two-meson scattering states can contribute to the $\Pi^{\text{Phen}}(q)$ in Eq. (\ref{corphen0}). Since these contributions are small, they can be neglected \cite{Lee:2004xk,Wang:2015nwa,Sundu:2018nxt,Agaev:2018vag,Wang:2020cme,Wang:2020eew}.

In the second step, correlation function in Eq. (\ref{corfunc}) is calculated via the Operator Product Expansion (OPE) formulated in Ref. \cite{Wilson:1969zs}. The OPE representation of the correlation function  $\Pi(q)$ can be obtained by inserting interpolating current into the correlation function and contracting heavy and light quark fields. This contraction is called Wick contraction. The result is composed of heavy and light quark propagators:
\begin{eqnarray}
\Pi^{\text{OPE}}(q)&=&  \int d^4x e^{iqx}  \text{Tr} [ S^{a^\prime a} q_(-x) \gamma_{\mu} S^{a a^\prime} c(x) \gamma_{\nu}] \nonumber  \\ 
&\times& \text{Tr} [S^{b^\prime b}_c(-x) \gamma_{\mu} S^{b b^\prime }_q(x) \gamma_{\nu} ] \label{qcdside}.
\end{eqnarray}
Here $S^{ab}_q$ and $S^{ab}_c$ are the light and heavy quark propagators, respectively. The light quark propagator reads as 
\begin{eqnarray}
 S^{ab}_q(x)&=&i \delta_{ab} \frac{\xslash x}{2\pi^2 x^4}- \delta_{ab} \frac{\langle \bar{q}q \rangle}{12} + i\delta_{ab} \frac{\xslash m_q \langle \bar{q}q \rangle}{48} \nonumber \\ 
 &-& \delta_{ab} \frac{x^2}{192} \langle \bar{q} g_s \sigma G q \rangle + i \delta_{ab} \frac{x^2 \xslash m_q }{1152}  \langle \bar{q} g_s \sigma G q \rangle \nonumber \\
 &-& ig_s \frac{G^{\alpha \beta}_{ab}}{32 \pi^2 x^2} [\xslash \sigma_{\alpha \beta}+ \sigma_{\alpha \beta} \xslash]- i\delta_{ab} \frac{x^2 \xslash g_s^2 \langle \bar{q}q \rangle^2}{7776} \nonumber \\
 &-& \delta_{ab} x^4 \frac{\langle \bar{q}q \rangle \langle g_s^2 G^2 \rangle}{27648} + \cdots. \label{light}
\end{eqnarray}
In the light quark propagator we set $m_q \to 0$ for $q=u,d$. For the heavy quark propagator $S^{ab}_c$, we employ the following expression
\begin{eqnarray}
S^{ab}_c(x) &=& i \int \frac{d^4k}{(2\pi)^4} e^{-ikx} [  \frac{\delta_{ab} (\kslash + m_c)}{k^2-m_c^2} \nonumber \\ 
&-& \frac{g_sG^{\alpha \beta}_{ab}}{4}\frac{\sigma_{\alpha \beta} (\kslash + m_c) +(\kslash + m_c)\sigma_{\alpha \beta}}{(k^2-m_c^2)^2} \nonumber \\
&+& \frac{g_s^2G^2 }{12} \delta_{ab} m_c \frac{k^2 +m_c \kslash}{(k^2-m_c^2)^4} + \cdots.
] \label{heavy}
\end{eqnarray} 
In Eqs. (\ref{light}) and (\ref{heavy}), we use the notations $G^{\alpha \beta}_{ab}=G^{\alpha \beta}_{A}t^A_{\alpha \beta}, ~ G^2=G^{A}_{\alpha \beta}G^{A}_{\alpha \beta}$, where $a,b=1,2,3$ and $A,B,C=1,2,\cdots,8$ are color indices. $t^A=\lambda^A/2$, and $\lambda^A$ are the Gell-Mann matrices. In the nonperturbative terms the gluon field strength tensor $G^{A}_{\alpha \beta}=G^{A}_{\alpha \beta}(0)$ is fixed at $x=0$.

The sum rules for mass $m_H$ and decay constant $f_H$ can be obtained via equating the invariant amplitudes $\Pi^{\text{Phen}}(q)$ and $\Pi^{\text{OPE}}(q)$; $\Pi^{\text{Phen}}(q)=\Pi^{\text{OPE}}(q)$. Such a direct matching is weakened due to the OPE is valid at large $q^2$ and phenomenological side is valid at small $q^2$. Furthermore, contributions of higher states and continuum still exist. Borel transformation should be applied in both sides of this equality in order to suppress contributions of higher states and continuum. After this operation, there is still a remnant in the sum rules. Using quark-hadron duality assumption, one subtracts higher states and continuum terms from the phenomenological side of the matching (the $\cdots$ terms in Eq. \ref{corphen1}). After these standard manipulations of the QCDSR method, the QPE side of the obtained sum rule takes the following form
\begin{equation}
\tilde{\Pi}^{\text{QCD}}=\int_{\mathcal{M}^2}^{s_0} \rho(s) e^{-s/M^2} ds, \label{boreledope}
\end{equation}
where $\mathcal{M}=(2m_c+2m_s)$ for $D_s^+ D_s^-$, $\mathcal{M}=(2m_c)$ for $D \bar D$ (we neglect the mass of $u$ and $d$ quarks), $s_0$ is the continuum threshold which is the energy characterizing beginning of the continuum, $M^2$ is Borel parameter and $\rho(s)$ is the spectral density obtained from the imaginary part of the correlation function, $\rho(s)=\text{Im}[\Pi(s)]/\pi$. We have taken into  account vacuum condensates up to dimension eight. The expressions for $\rho(s)$ are rather lengthy, so we refrain to provide it here. Interested readers can request the spectral densities from the author.

The phenomenological side of the sum rule has the following form
\begin{equation}
\tilde{\Pi}^{\text{Phen}}=f_H^2 m_H^2 e^{-\frac{m_H^2}{M^2}}. \label{boreledphen}
\end{equation}
Equating Eqs. (\ref{boreledope}) and (\ref{boreledphen}), one can obtain sum rules for mass and decay constant. The mass formula reads as
\begin{equation}
m_H^2 (s_0, M^2)=\frac{\tilde{\Pi}^{\prime \text{QCD}}}{\tilde{\Pi}^{\text{QCD}}} 
\end{equation}
where $\tilde{\Pi}^{\prime \text{QCD}}=\frac{d}{d(-\frac{1}{M^2})} \tilde{\Pi}^{\text{QCD}}$, and decay constant formula reads as
\begin{equation}
f_H^2 (s_0, M^2)  = \frac{e^{m_H^2/M^2}}{m_H^2} \tilde{\Pi}^{\text{QCD}}.
\end{equation}

\section{\label{sec:level3}Spectroscopic Parameters of  $D \bar D$ and $D_s^+ D_s^-$ States}
There are two parameter groups for a QCDSR calculation; QCD input parameters and resulting sum rule parameters of the model. The first group consists of quark masses and QCD condensates which are given as
\begin{eqnarray}
&&\langle \overline{q}q\rangle =-(0.24\pm 0.01)^{3}~\mathrm{GeV}^{3},\
\langle \overline{s}s\rangle =(0.8\pm 0.1)\langle \overline{q}q\rangle,
\notag \\
&&\langle \overline{q}g_{s}\sigma Gq\rangle =m_{0}^{2}\langle \overline{q}%
q\rangle,\ \langle \overline{s}g_{s}\sigma Gs\rangle =m_{0}^{2}\langle
\overline{s}s\rangle,  \notag \\
&&m_{0}^{2}=(0.8\pm 0.2)~\mathrm{GeV}^{2}  \notag \\
&&\langle \frac{\alpha _{s}G^{2}}{\pi }\rangle =(0.012\pm 0.004)~\mathrm{GeV}%
^{4},  \notag \\
&&m_{s}=93_{-5}^{+11}~\mathrm{MeV},\ m_{c}=1.27\pm 0.2~\mathrm{GeV}.
\label{eq:QCDParameters}
\end{eqnarray}
The second group contains the continuum threshold parameter $s_0$ which arises after continuum subtraction via using quark-hadron duality and the $M^2$ Borel parameter which arises after Borel transformation. These are auxiliary parameters and any quantity which is extracted from QCDSR calculations should be independent or have a mild dependence with respect to variation of these parameters. Notwithstanding, these auxiliary parameters should meet some criteria of the QCDSR method. Since there is no single value for $s_0$ and $M^2$, a suitable working regions of these parameters must be found in order to get reliable predictions. Finding proper choice of continuum threshold parameter $s_0$ and Borel parameter $M^2$ is an important task in the sum rule analysis. 

The region for Borel parameter $M^2$ is determined by OPE convergence and suppression of the contributions arising from the higher and continuum states. Finding upper and lower limits of Borel parameter requires additional elaboration of the sum rules. Pole contribution (PC), 
\begin{equation}
\text{PC}=\frac{\Pi^{\text{OPE}}(s_0,M^2)}{\Pi^{\text{OPE}}(s_0=\infty,M^2)}, \label{PC}
\end{equation}
and convergence of the operator product expansion
\begin{equation}
R(M^2)=\frac{\Pi^{\text{OPE,DimN}}(s_0,M^2)}{\Pi^{\text{OPE}}(s_0=\infty,M^2)}. \label{OPEConverg}
\end{equation}
will help to determine upper and lower limits of the Borel parameter. In Eq. \ref{OPEConverg}, $\Pi^{\text{OPE,DimN}}(s_0,M^2)$ is the expression used to estimate convergence of OPE. In QCD sum rule approach, $\mathcal{O}(\alpha_s)$ corrections to leading and next-to-leading order terms in $\Pi^{\text{OPE}}(s_0,M^2)$ maybe comparable above Dim 8 dimensional contributions. Due to the interpolating current we used, contribution of Dim8 disappears due to the gamma matrix algebra. Therefore we use $\Pi^{\text{Dim7}}(s_0,M^2)$ to look for the convergence of OPE. In this way, we can avoid uncertainties connected to $\mathcal{O}(\alpha_s)$ corrections.

For the upper limit, PC should reside fairly in the correlation function while for the lower limit PC must be dominant. This is due to the fact that after Borel transformation, an exponential term emerges in both side of the sum rule, as can be seen from Eqs. \ref{boreledope} and \ref{boreledphen}. The integrals receive main contribution at $s \simeq M^2$. For lower $M^2$ values, the integral (Eq. \ref{boreledope}) receives contributions from lower $s$ values and for higher $M^2$ values these contributions decrease. $R(M^2)$ can be used to determine the lower limit of Borel parameter. Between upper and lower limits of $M^2$ a balance can be maintained where dominance of single resonance contribution and convergence of the series take place.

The continuum threshold $s_0$ eliminates contributions from the continuum states and higher resonances, and isolates the ground state in the correlation function. $s_0$ is the place where the continuum states and higher resonances begin to contribute to the correlation function. Therefore $s_0$ specifies the threshold in the integral which needs to be carried out. Since in the sum rule calculations ground states are being handled, continuum threshold $s_0$ starts with the first excited state. The energy that needs to excite the ground state is $\delta = \sqrt{s_0}-m$ where $m$ is the mass of the ground state. In conventional mesons and baryons, parameters of first excited states are known experimentally or have a theoretical consensus. Based on this knowledge, it is quite possible to estimate $s_0$ value. For experimentally known mesons and baryons $\delta$ varies in the range of $0.3 ~\text{GeV} \leq \delta \leq 0.8 ~\text{GeV}$. In the case of multiquark hadrons (exotic states) there can be lack of relevant information for excited states and ground states. The determination of $s_0$ can be obtained via aforementioned limits on PC and the convergence of OPE. The region for continuum threshold $s_0$ can be chosen as possible as small to provide a reliable Borel region and meanwhile to obtain a maximum for PC. With all these considerations it is possible to examine obtained sum rules numerically. Our analysis gives

\begin{eqnarray}
M^{2}\in [2.0,2.4]~\mathrm{GeV}^{2},\ s_{0}\in [17,19]~\mathrm{GeV}^{2},  \label{worechi} \\
M^{2}\in [2.3,2.7]~\mathrm{GeV}^{2},\ s_{0}\in [19,21]~\mathrm{GeV}^{2}, \label{worex(3960)}
\end{eqnarray}%
for $D \bar D$ and $D_s^+ D_s^-$, respectively. The working regions of Eqs. (\ref{worechi}) and (\ref{worex(3960)}) meet all the restrictions mentioned before. A further discussion should be done for the continuum threshold $s_0$ value. Considering $D^{*\pm}(2010)$ state for the first excited state of $D$, the mass \cite{Workman:2022ynf} square of two-$D^{*\pm}(2010)$ meson gives $(2m_{D^{*\pm}})^2 \simeq 16.16 ~\mathrm{GeV}^{2}$. On the otherside, considering $D^{*0}(2007)$ state for the first excited state of $D$,  the mass \cite{Workman:2022ynf} square of two-$D^{*0}(2007)$ meson gives $(2m_{D^{*0}})^2 \simeq 16.10 ~\mathrm{GeV}^{2}$. By same token, considering $D_s^{*\pm}$ state as the first excited state of $D_s^{\pm}$, the mass \cite{Workman:2022ynf} square of two-$D_s^{*\pm}$ meson gives $(2m_{D_s^{*\pm}})^2 \simeq 17.84 ~\mathrm{GeV}^{2}$. The obtained continuum threshold $s_0$ values are reasonable to excite $D \bar D$ and $D_s^+ D_s^-$ to their first excited states. 

Working regions of Borel parameter $M^2$ and continuum threshold $s_0$ give the pole contributions as
\begin{eqnarray}
D \bar D & : & 0.32 \leq \text{PC} \leq 0.62, \\
D_s^+ D_s^- & : & 0.29 \leq \text{PC} \leq 0.73, 
\end{eqnarray}
for upper and lower limits of Borel parameter $M^2$, respectively. In a standard QCDSR analysis, the minimum value for PC of four-quark systems is approximately 0.15-0.2 \cite{Agaev:2020zad}. At the upper limit of Borel parameter $M^2$, PC should exceed this range. After providing PC in an appropriate range, one can extract the mass and decay constant values for the corresponding state. Our results for masses and decay constants of $D \bar D$ and $D_s^+ D_s^-$ states read as
\begin{eqnarray}
M_{D \bar D} &=& 3795^{+85}_{-82} ~\mathrm{MeV}, \nonumber \\
 f_{D \bar D} &=& 1.70^{+0.33}_{-0.29} \times 10^{-2}  ~\mathrm{GeV}^5, \label{result1}
\end{eqnarray}
and
\begin{eqnarray}
M_{D_s^+ D_s^-} &=& 3983^{+93}_{-88} ~\mathrm{MeV}, \nonumber \\
f_{D_s^+ D_s^-} &=& 2.52^{+0.64}_{-0.54} \times 10^{-2}  ~\mathrm{GeV}^5. \label{result2}
\end{eqnarray}
In Figures \ref{fig:Mass1}, \ref{fig:decayconstant1}, \ref{fig:Mass2} and \ref{fig:decayconstant2} we plot the mass and decay constant of $D \bar D$ and $D_s^+ D_s^-$ states as functions of $M^2$ and $s_0$, respectively.

\begin{widetext}

\begin{figure}[h]
\begin{center}
\includegraphics[totalheight=6cm,width=8cm]{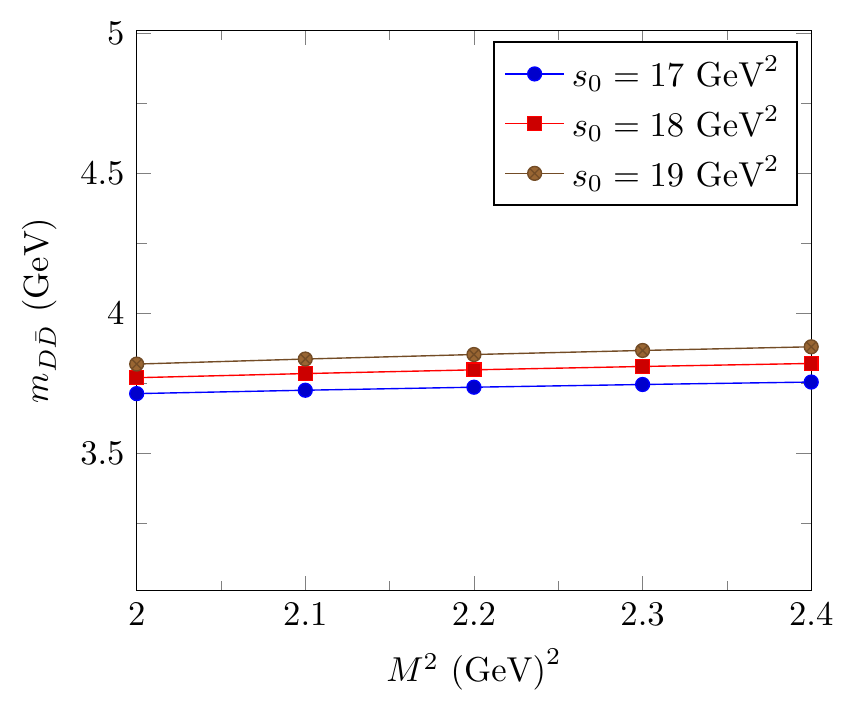}
\includegraphics[totalheight=6cm,width=8cm]{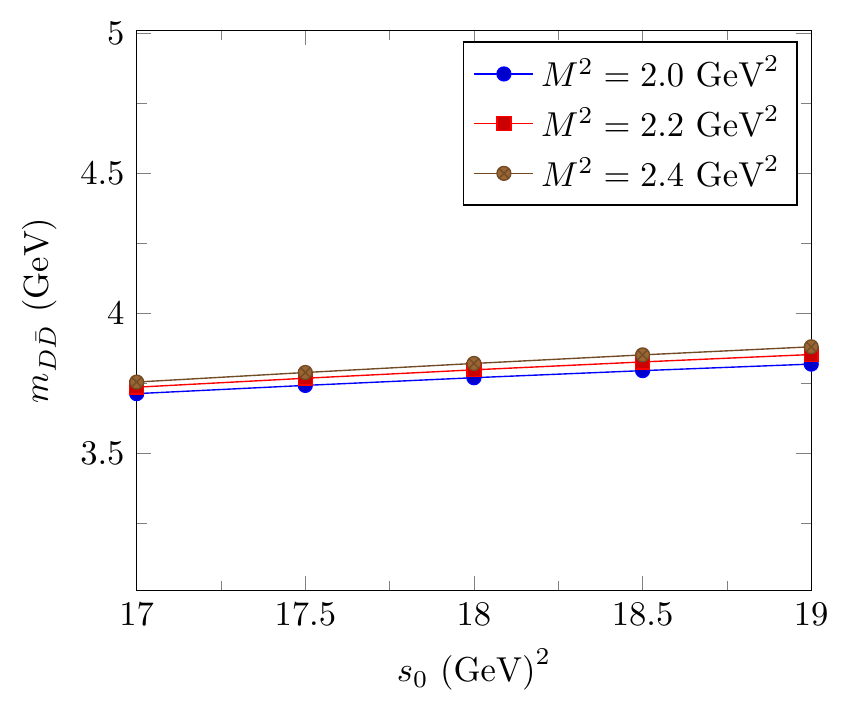}
\end{center}
\caption{The mass of the  $D \bar D$ state as a function of $M^{2}$ at fixed $s_{0} $ (left panel), and as a function of $s_0$ at fixed $M^2$ (right panel).}
\label{fig:Mass1}
\end{figure}
\begin{figure}[h]
\begin{center}
\includegraphics[totalheight=6cm,width=8cm]{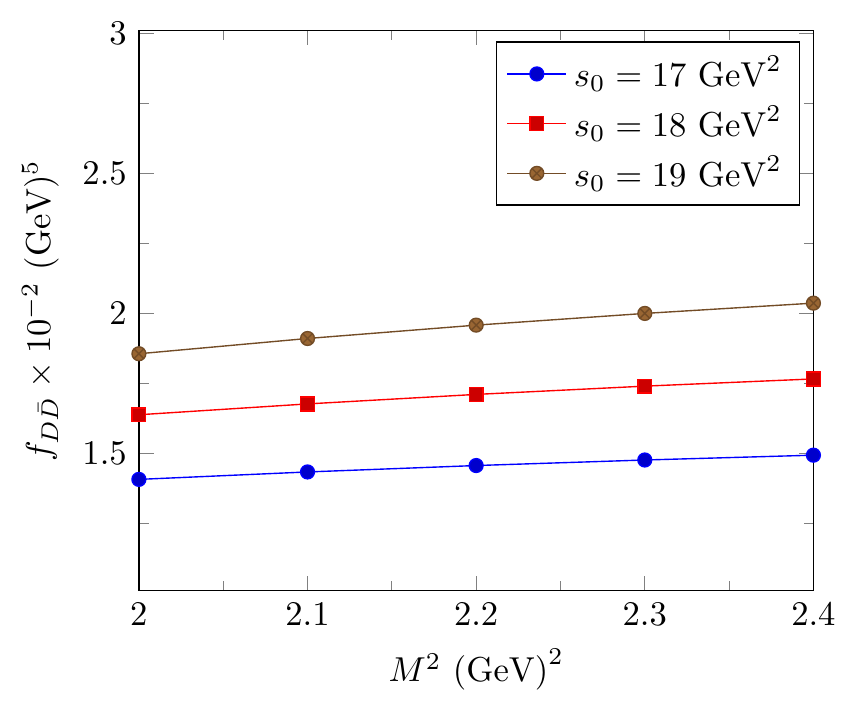}\,\, %
\includegraphics[totalheight=6cm,width=8cm]{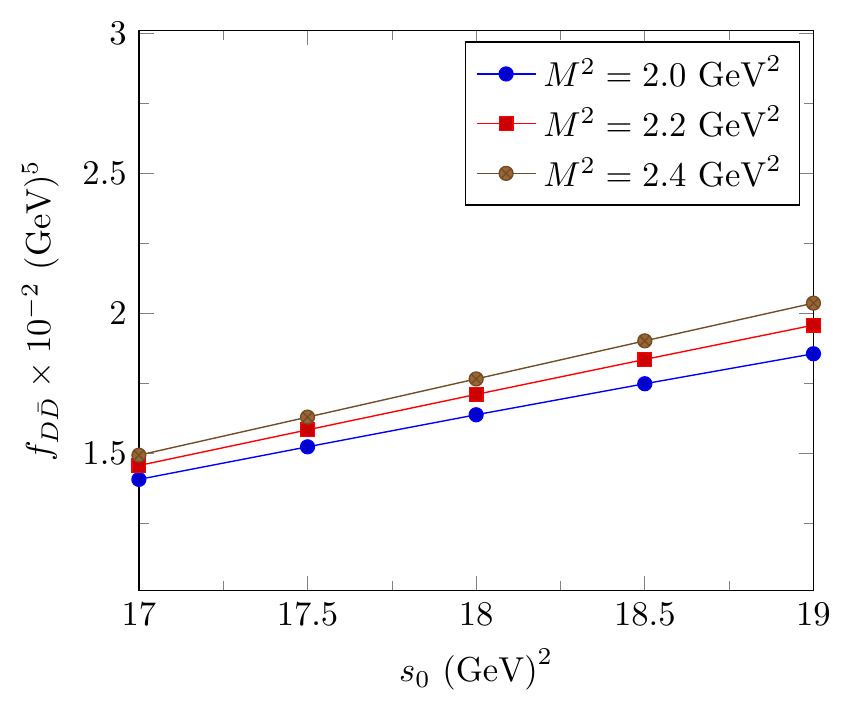}
\end{center}
\caption{Same as in Fig. \ref{fig:Mass1}, but for the coupling $f$ of the $D \bar D$ state.}
\label{fig:decayconstant1}
\end{figure}

\begin{figure}[h]
\begin{center}
\includegraphics[totalheight=6cm,width=8cm]{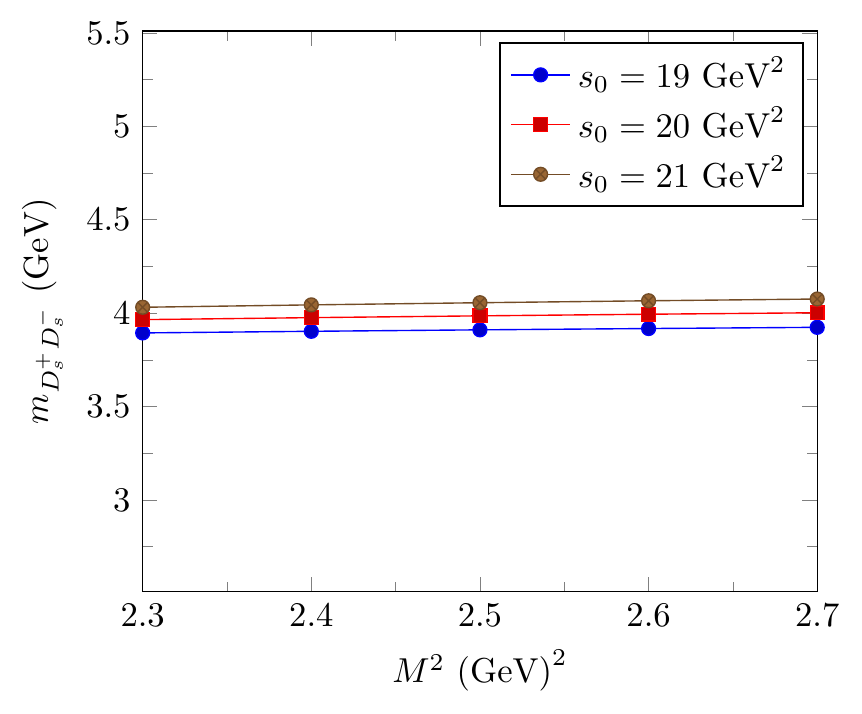}
\includegraphics[totalheight=6cm,width=8cm]{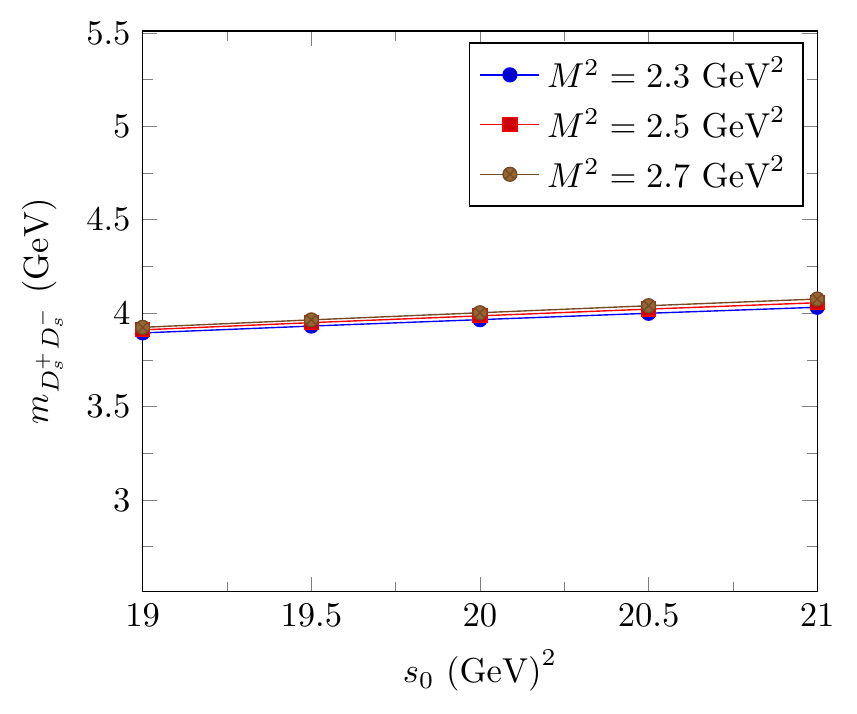}
\end{center}
\caption{The mass of the  $D_s^+ D_s^-$ state as a function of $M^{2}$ at fixed $s_{0} $ (left panel), and as a function of $s_0$ at fixed $M^2$ (right panel).}
\label{fig:Mass2}
\end{figure}
\begin{figure}[h]
\begin{center}
\includegraphics[totalheight=6cm,width=8cm]{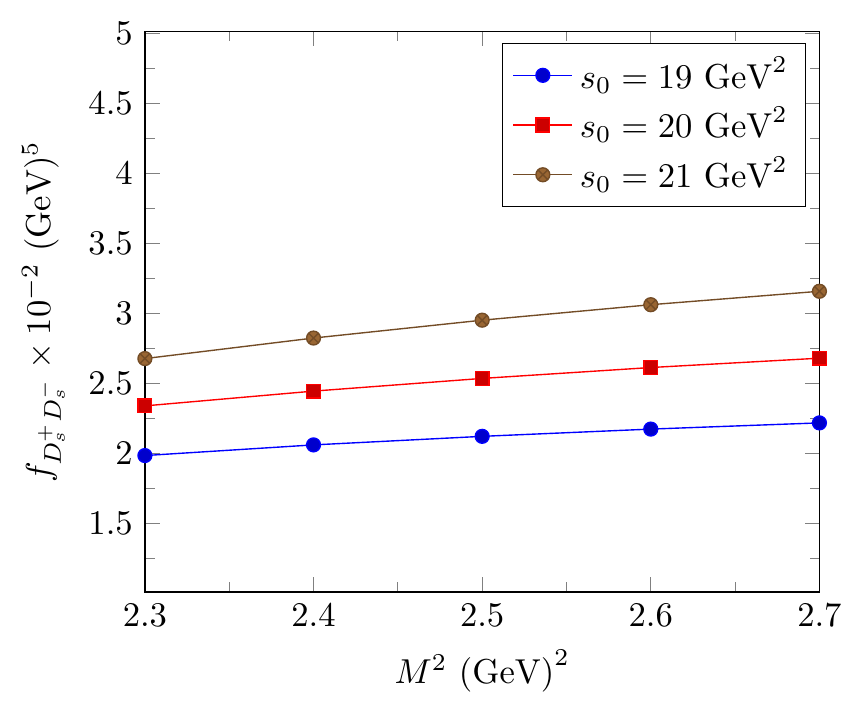}\,\, %
\includegraphics[totalheight=6cm,width=8cm]{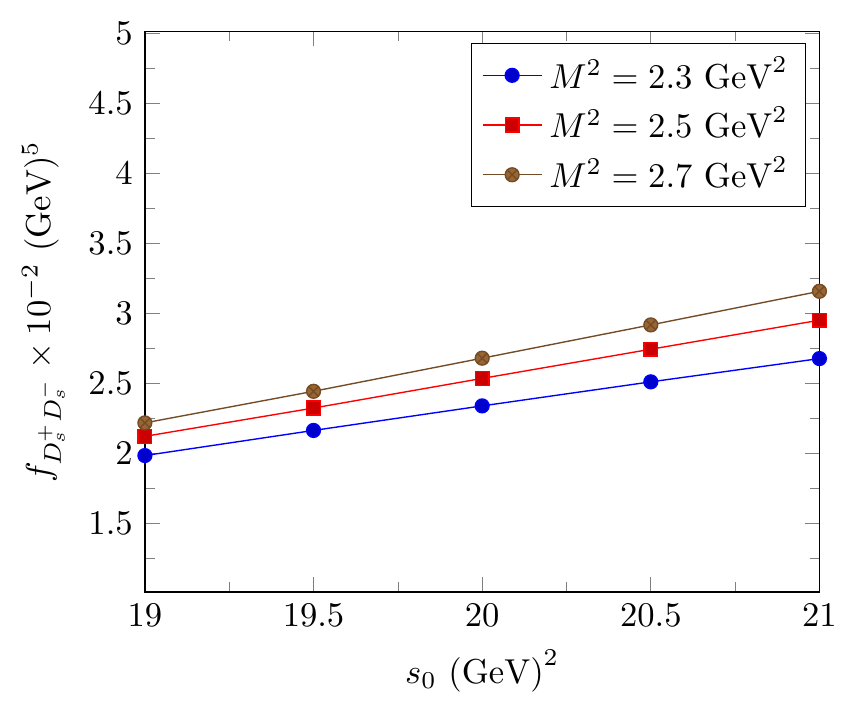}
\end{center}
\caption{Same as in Fig. \ref{fig:Mass2}, but for the coupling $f$ of the $D_s^+ D_s^-$  state.}
\label{fig:decayconstant2}
\end{figure}

\end{widetext}

As can be seen from Figs. \ref{fig:Mass1} and \ref{fig:Mass2}, the mass values are very stable with respect to changes in Borel parameter $M^2$ and continuum threshold $s_0$. Figs. \ref{fig:decayconstant1} and \ref{fig:decayconstant2} have a residual dependence on the Borel parameter $M^2$ and continuum threshold $s_0$. This dependence is an essential part of the theoretical errors in QCD sum rule calculations. The reason for this situation could be as a result of mass and decay constant extractions from the QCD sum rules. The mass of the state is determined as $m^2=\frac{\Pi^\prime(s_0,M^2)}{\Pi(s_0,M^2)}$. The spectral densities in both numerator and denominator  could smooth variations of $M^2$ and $s_0$. On the other side, the decay constant is determined as $f^2=\frac{e^{m^2/M^2}}{m^2} \Pi(s_0,M^2)$ where the numerical value can change with respect to changes in $M^2$ and $s_0$ since such a ratio (as in the case of mass) of correlation functions is absent. Theoretical uncertainties of the extracted masses are 2.1 \% for $D \bar D$ state and  2.1 \% for $D_s^+ D_s^-$  state. In the decay constant values, theoretical uncertainties are 18 \% for $D \bar D$ state and 23 \% for $D_s^+ D_s^-$  state. These theoretical uncertainties are well below the accepted limits in QCDSR calculations.

The extracted mass $M_{D_s^+ D_s^-}=3983^{+93}_{-88} ~\mathrm{MeV}$ for $D_s^+ D_s^-$ state with the $J^{PC}=0^{++}$ quantum number agree well with the experimental mass $M=3955 \pm 6 \pm 11 ~ \text{MeV}$ of $X(3960)$. Our prediction supports $X(3960)$ to be a $D_s^+ D_s^-$ molecular state with $J^{PC}=0^{++}$ quantum number. A recent QCDSR study obtained mass of $D_s^+ D_s^-$ state as $M= 3.98 \pm 0.10 ~\mathrm{GeV} $ and decay constant $f= 2.36^{+0.45}_{-0.45} \times 10^{-2}  ~\mathrm{GeV}^5$, with $J^{PC}=0^{++}$ in the molecular picture \cite{Xin:2022bzt}. Our results agree well with their predictions. 

\begin{figure}[h!]
\centering
\includegraphics[width=3.4in]{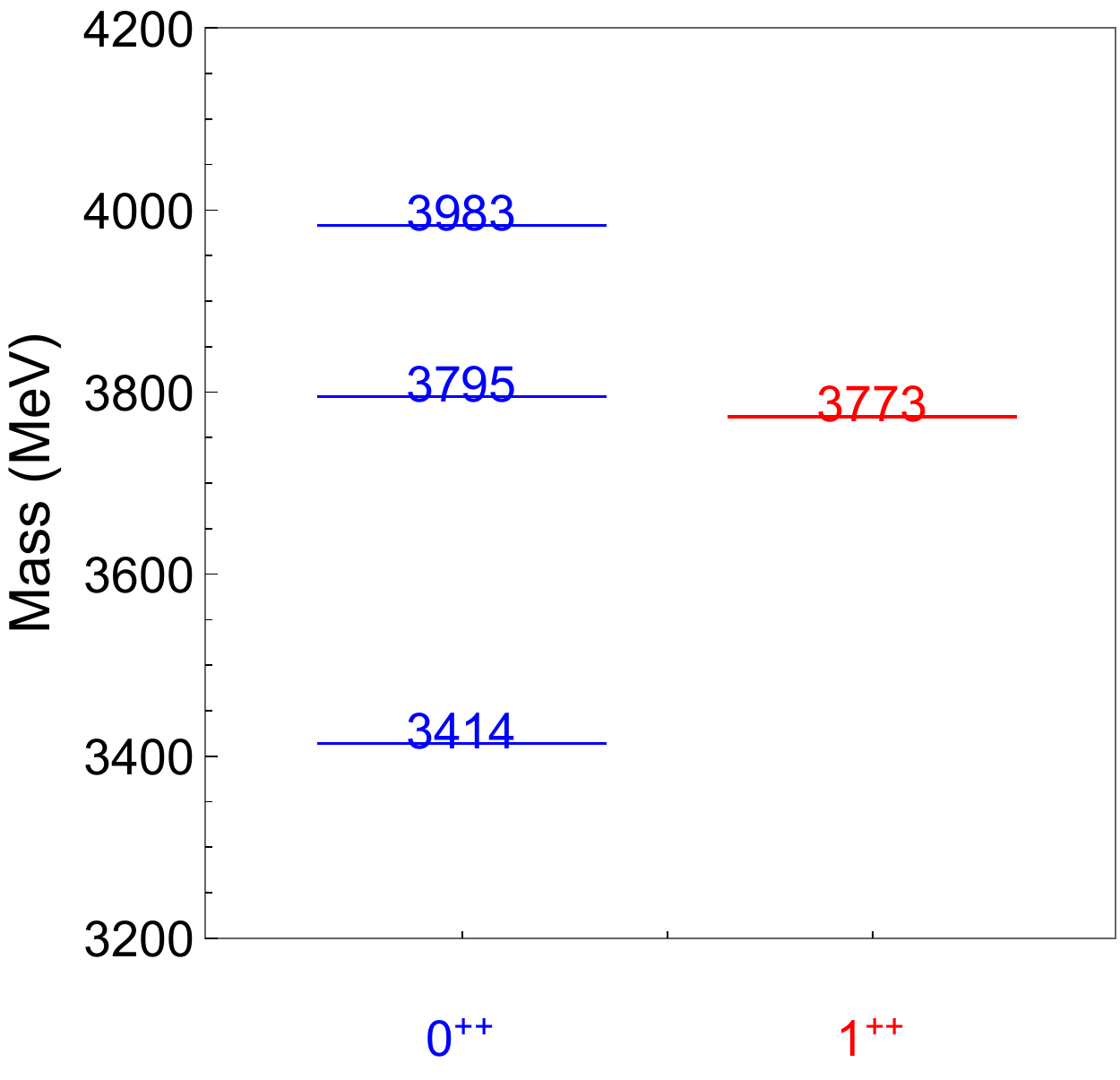}
\caption{\label{fig:location} Predicted mass values of $D \bar D$ and $D_s^+ D_s^-$ states, and the lowest $\chi_{c0}(1P)$ state together with $\psi(3770)$. The lowest state in $J^{PC}=0^{++}$ is $\chi_{c0}(1P)$ and the state in $J^{PC}=1^{++}$ is $\psi(3770)$. 3795 and 3983 refer to mass values of $D \bar D$ and $D_s^+ D_s^-$ states in $J^{PC}=0^{++}$, respectively. } 
\end{figure}

 The extracted mass for $D \bar D$ state with the $J^{PC}=0^{++}$ quantum number is $M_{D \bar D}=3795^{+85}_{-82} ~\mathrm{MeV}$. Ref. \cite{Xin:2022bzt} obtained mass and decay constant of $D \bar D$ state as $M=3.74 \pm 0.09 ~\mathrm{GeV}$ and $f= 1.61^{+0.23}_{-0.23} \times 10^{-2}  ~\mathrm{GeV}^5$. Our results agree well with their predictions. The existence of such a bound state near the $D \bar D$ threshold was demonstrated by analysis of $ \gamma \gamma \to D \bar D$ \cite{Wang:2020elp,Deineka:2021aeu}. The extracted mass of $D \bar D$ state lies nearby by the $\psi(3770)$ with mass $M=3773.7 \pm 0.4 ~\mathrm{MeV}$. The quantum number of $\psi(3770)$ is $J^{PC}=1^{++}$. Conventional $\chi_{c0}(1P)$ state has a mass of $M=3414.71 \pm 0.30 ~\mathrm{MeV}$ with $J^{PC}=0^{++}$ quantum number. The locations of these states can be seen in Fig. \ref{fig:location}. The $D \bar D$ state has no overlaps with these states. Therefore possible observation of $D \bar D$ state is important in terms of establishing the lowest four-quark states in charmonium sector. 

\section{\label{sec:level4}Epilogue}
In this present study, we considered $D \bar D$ and $D_s^+ D_s^-$ states with $J^{PC}=0^{++}$ quantum number in molecular picture. We extracted mass and decay constant (residue) values by using two-point QCDSR formalism. We carried out the OPE up to the vacuum condensates of dimension eight. Establishing pole contribution within the limits of QCDSR, we determined working regions of continuum threshold $s_0$ and Borel parameter $M^2$. The extracted mass and decay constant values of $D \bar D$ and $D_s^+ D_s^-$ states read as $M_{D \bar D} = 3795^{+85}_{-82} ~\mathrm{MeV} $, $f_{D \bar D} = 1.70^{+0.33}_{-0.29} \times 10^{-2}  ~\mathrm{GeV}^5$, and $M_{D_s^+ D_s^-} = 3983^{+93}_{-88} ~\mathrm{MeV}$, $f_{D_s^+ D_s^-} = 2.52^{+0.64}_{-0.54} \times 10^{-2}  ~\mathrm{GeV}^5$, respectively. The predicted mass for $D_s^+ D_s^-$ state agrees well with the experimental value $M=3955 \pm 6 \pm 11 ~ \text{MeV}$ of the very recent LHCb observation of $X(3960)$. Our result for $D_s^+ D_s^-$ state supports the quantum number assignment and molecular picture of $X(3960)$. On the other side, the possible observation of  $D \bar D$ state may pave the fay for the lowest four-quark state in charmonium sector. We hope our outcomes may be helpful for future theoretical studies and experimental endeavours.  

\begin{acknowledgments}
The author of this work is supported by The Scientific and Technological Research Council of Türkiye (TUBITAK) in the framework of BIDEB-2219 International Postdoctoral Research Fellowship Program.
\end{acknowledgments}

\bibliography{ccqq}

\begin{thebibliography}{48}%
\makeatletter
\providecommand \@ifxundefined [1]{%
 \@ifx{#1\undefined}
}%
\providecommand \@ifnum [1]{%
 \ifnum #1\expandafter \@firstoftwo
 \else \expandafter \@secondoftwo
 \fi
}%
\providecommand \@ifx [1]{%
 \ifx #1\expandafter \@firstoftwo
 \else \expandafter \@secondoftwo
 \fi
}%
\providecommand \natexlab [1]{#1}%
\providecommand \enquote  [1]{``#1''}%
\providecommand \bibnamefont  [1]{#1}%
\providecommand \bibfnamefont [1]{#1}%
\providecommand \citenamefont [1]{#1}%
\providecommand \href@noop [0]{\@secondoftwo}%
\providecommand \href [0]{\begingroup \@sanitize@url \@href}%
\providecommand \@href[1]{\@@startlink{#1}\@@href}%
\providecommand \@@href[1]{\endgroup#1\@@endlink}%
\providecommand \@sanitize@url [0]{\catcode `\\12\catcode `\$12\catcode
  `\&12\catcode `\#12\catcode `\^12\catcode `\_12\catcode `\%12\relax}%
\providecommand \@@startlink[1]{}%
\providecommand \@@endlink[0]{}%
\providecommand \url  [0]{\begingroup\@sanitize@url \@url }%
\providecommand \@url [1]{\endgroup\@href {#1}{\urlprefix }}%
\providecommand \urlprefix  [0]{URL }%
\providecommand \Eprint [0]{\href }%
\providecommand \doibase [0]{http://dx.doi.org/}%
\providecommand \selectlanguage [0]{\@gobble}%
\providecommand \bibinfo  [0]{\@secondoftwo}%
\providecommand \bibfield  [0]{\@secondoftwo}%
\providecommand \translation [1]{[#1]}%
\providecommand \BibitemOpen [0]{}%
\providecommand \bibitemStop [0]{}%
\providecommand \bibitemNoStop [0]{.\EOS\space}%
\providecommand \EOS [0]{\spacefactor3000\relax}%
\providecommand \BibitemShut  [1]{\csname bibitem#1\endcsname}%
\let\auto@bib@innerbib\@empty
%</preamble>
\bibitem [{\citenamefont {Gell-Mann}(1964)}]{Gell-Mann:1964ewy}%
  \BibitemOpen
  \bibfield  {author} {\bibinfo {author} {\bibfnamefont {M.}~\bibnamefont
  {Gell-Mann}},\ }\href {\doibase 10.1016/S0031-9163(64)92001-3} {\bibfield
  {journal} {\bibinfo  {journal} {Phys. Lett.}\ }\textbf {\bibinfo {volume}
  {8}},\ \bibinfo {pages} {214} (\bibinfo {year} {1964})}\BibitemShut {NoStop}%
\bibitem [{\citenamefont {Zweig}(1964)}]{Zweig:1964ruk}%
  \BibitemOpen
  \bibfield  {author} {\bibinfo {author} {\bibfnamefont {G.}~\bibnamefont
  {Zweig}},\ }\href@noop {} {\  (\bibinfo {year} {1964})}\BibitemShut {NoStop}%
\bibitem [{\citenamefont {Choi}\ \emph {et~al.}(2003)\citenamefont {Choi} \emph
  {et~al.}}]{Belle:2003nnu}%
  \BibitemOpen
  \bibfield  {author} {\bibinfo {author} {\bibfnamefont {S.~K.}\ \bibnamefont
  {Choi}} \emph {et~al.} (\bibinfo {collaboration} {Belle}),\ }\href {\doibase
  10.1103/PhysRevLett.91.262001} {\bibfield  {journal} {\bibinfo  {journal}
  {Phys. Rev. Lett.}\ }\textbf {\bibinfo {volume} {91}},\ \bibinfo {pages}
  {262001} (\bibinfo {year} {2003})},\ \Eprint
  {http://arxiv.org/abs/hep-ex/0309032} {arXiv:hep-ex/0309032} \BibitemShut
  {NoStop}%
\bibitem [{\citenamefont {Bondar}\ \emph {et~al.}(2012)\citenamefont {Bondar}
  \emph {et~al.}}]{Belle:2011aa}%
  \BibitemOpen
  \bibfield  {author} {\bibinfo {author} {\bibfnamefont {A.}~\bibnamefont
  {Bondar}} \emph {et~al.} (\bibinfo {collaboration} {Belle}),\ }\href
  {\doibase 10.1103/PhysRevLett.108.122001} {\bibfield  {journal} {\bibinfo
  {journal} {Phys. Rev. Lett.}\ }\textbf {\bibinfo {volume} {108}},\ \bibinfo
  {pages} {122001} (\bibinfo {year} {2012})},\ \Eprint
  {http://arxiv.org/abs/1110.2251} {arXiv:1110.2251 [hep-ex]} \BibitemShut
  {NoStop}%
\bibitem [{\citenamefont {Ablikim}\ \emph {et~al.}(2013)\citenamefont {Ablikim}
  \emph {et~al.}}]{BESIII:2013ouc}%
  \BibitemOpen
  \bibfield  {author} {\bibinfo {author} {\bibfnamefont {M.}~\bibnamefont
  {Ablikim}} \emph {et~al.} (\bibinfo {collaboration} {BESIII}),\ }\href
  {\doibase 10.1103/PhysRevLett.111.242001} {\bibfield  {journal} {\bibinfo
  {journal} {Phys. Rev. Lett.}\ }\textbf {\bibinfo {volume} {111}},\ \bibinfo
  {pages} {242001} (\bibinfo {year} {2013})},\ \Eprint
  {http://arxiv.org/abs/1309.1896} {arXiv:1309.1896 [hep-ex]} \BibitemShut
  {NoStop}%
\bibitem [{\citenamefont {Ablikim}\ \emph {et~al.}(2014)\citenamefont {Ablikim}
  \emph {et~al.}}]{BESIII:2013mhi}%
  \BibitemOpen
  \bibfield  {author} {\bibinfo {author} {\bibfnamefont {M.}~\bibnamefont
  {Ablikim}} \emph {et~al.} (\bibinfo {collaboration} {BESIII}),\ }\href
  {\doibase 10.1103/PhysRevLett.112.132001} {\bibfield  {journal} {\bibinfo
  {journal} {Phys. Rev. Lett.}\ }\textbf {\bibinfo {volume} {112}},\ \bibinfo
  {pages} {132001} (\bibinfo {year} {2014})},\ \Eprint
  {http://arxiv.org/abs/1308.2760} {arXiv:1308.2760 [hep-ex]} \BibitemShut
  {NoStop}%
\bibitem [{\citenamefont {Aaij}\ \emph {et~al.}(2015)\citenamefont {Aaij} \emph
  {et~al.}}]{LHCb:2015yax}%
  \BibitemOpen
  \bibfield  {author} {\bibinfo {author} {\bibfnamefont {R.}~\bibnamefont
  {Aaij}} \emph {et~al.} (\bibinfo {collaboration} {LHCb}),\ }\href {\doibase
  10.1103/PhysRevLett.115.072001} {\bibfield  {journal} {\bibinfo  {journal}
  {Phys. Rev. Lett.}\ }\textbf {\bibinfo {volume} {115}},\ \bibinfo {pages}
  {072001} (\bibinfo {year} {2015})},\ \Eprint
  {http://arxiv.org/abs/1507.03414} {arXiv:1507.03414 [hep-ex]} \BibitemShut
  {NoStop}%
\bibitem [{\citenamefont {Aaij}\ \emph {et~al.}(2019)\citenamefont {Aaij} \emph
  {et~al.}}]{LHCb:2019kea}%
  \BibitemOpen
  \bibfield  {author} {\bibinfo {author} {\bibfnamefont {R.}~\bibnamefont
  {Aaij}} \emph {et~al.} (\bibinfo {collaboration} {LHCb}),\ }\href {\doibase
  10.1103/PhysRevLett.122.222001} {\bibfield  {journal} {\bibinfo  {journal}
  {Phys. Rev. Lett.}\ }\textbf {\bibinfo {volume} {122}},\ \bibinfo {pages}
  {222001} (\bibinfo {year} {2019})},\ \Eprint
  {http://arxiv.org/abs/1904.03947} {arXiv:1904.03947 [hep-ex]} \BibitemShut
  {NoStop}%
\bibitem [{\citenamefont {Ablikim}\ \emph {et~al.}(2021)\citenamefont {Ablikim}
  \emph {et~al.}}]{BESIII:2020qkh}%
  \BibitemOpen
  \bibfield  {author} {\bibinfo {author} {\bibfnamefont {M.}~\bibnamefont
  {Ablikim}} \emph {et~al.} (\bibinfo {collaboration} {BESIII}),\ }\href
  {\doibase 10.1103/PhysRevLett.126.102001} {\bibfield  {journal} {\bibinfo
  {journal} {Phys. Rev. Lett.}\ }\textbf {\bibinfo {volume} {126}},\ \bibinfo
  {pages} {102001} (\bibinfo {year} {2021})},\ \Eprint
  {http://arxiv.org/abs/2011.07855} {arXiv:2011.07855 [hep-ex]} \BibitemShut
  {NoStop}%
\bibitem [{\citenamefont {Abe}\ \emph {et~al.}(2005)\citenamefont {Abe} \emph
  {et~al.}}]{Belle:2004lle}%
  \BibitemOpen
  \bibfield  {author} {\bibinfo {author} {\bibfnamefont {K.}~\bibnamefont
  {Abe}} \emph {et~al.} (\bibinfo {collaboration} {Belle}),\ }\href {\doibase
  10.1103/PhysRevLett.94.182002} {\bibfield  {journal} {\bibinfo  {journal}
  {Phys. Rev. Lett.}\ }\textbf {\bibinfo {volume} {94}},\ \bibinfo {pages}
  {182002} (\bibinfo {year} {2005})},\ \Eprint
  {http://arxiv.org/abs/hep-ex/0408126} {arXiv:hep-ex/0408126} \BibitemShut
  {NoStop}%
\bibitem [{\citenamefont {Aubert}\ \emph {et~al.}(2008)\citenamefont {Aubert}
  \emph {et~al.}}]{BaBar:2007vxr}%
  \BibitemOpen
  \bibfield  {author} {\bibinfo {author} {\bibfnamefont {B.}~\bibnamefont
  {Aubert}} \emph {et~al.} (\bibinfo {collaboration} {BaBar}),\ }\href
  {\doibase 10.1103/PhysRevLett.101.082001} {\bibfield  {journal} {\bibinfo
  {journal} {Phys. Rev. Lett.}\ }\textbf {\bibinfo {volume} {101}},\ \bibinfo
  {pages} {082001} (\bibinfo {year} {2008})},\ \Eprint
  {http://arxiv.org/abs/0711.2047} {arXiv:0711.2047 [hep-ex]} \BibitemShut
  {NoStop}%
\bibitem [{\citenamefont {Lees}\ \emph {et~al.}(2012)\citenamefont {Lees} \emph
  {et~al.}}]{BaBar:2012nxg}%
  \BibitemOpen
  \bibfield  {author} {\bibinfo {author} {\bibfnamefont {J.~P.}\ \bibnamefont
  {Lees}} \emph {et~al.} (\bibinfo {collaboration} {BaBar}),\ }\href {\doibase
  10.1103/PhysRevD.86.072002} {\bibfield  {journal} {\bibinfo  {journal} {Phys.
  Rev. D}\ }\textbf {\bibinfo {volume} {86}},\ \bibinfo {pages} {072002}
  (\bibinfo {year} {2012})},\ \Eprint {http://arxiv.org/abs/1207.2651}
  {arXiv:1207.2651 [hep-ex]} \BibitemShut {NoStop}%
\bibitem [{\citenamefont {Aaij}\ \emph
  {et~al.}(2020{\natexlab{a}})\citenamefont {Aaij} \emph
  {et~al.}}]{LHCb:2020bls}%
  \BibitemOpen
  \bibfield  {author} {\bibinfo {author} {\bibfnamefont {R.}~\bibnamefont
  {Aaij}} \emph {et~al.} (\bibinfo {collaboration} {LHCb}),\ }\href {\doibase
  10.1103/PhysRevLett.125.242001} {\bibfield  {journal} {\bibinfo  {journal}
  {Phys. Rev. Lett.}\ }\textbf {\bibinfo {volume} {125}},\ \bibinfo {pages}
  {242001} (\bibinfo {year} {2020}{\natexlab{a}})},\ \Eprint
  {http://arxiv.org/abs/2009.00025} {arXiv:2009.00025 [hep-ex]} \BibitemShut
  {NoStop}%
\bibitem [{\citenamefont {Aaij}\ \emph
  {et~al.}(2020{\natexlab{b}})\citenamefont {Aaij} \emph
  {et~al.}}]{LHCb:2020pxc}%
  \BibitemOpen
  \bibfield  {author} {\bibinfo {author} {\bibfnamefont {R.}~\bibnamefont
  {Aaij}} \emph {et~al.} (\bibinfo {collaboration} {LHCb}),\ }\href {\doibase
  10.1103/PhysRevD.102.112003} {\bibfield  {journal} {\bibinfo  {journal}
  {Phys. Rev. D}\ }\textbf {\bibinfo {volume} {102}},\ \bibinfo {pages}
  {112003} (\bibinfo {year} {2020}{\natexlab{b}})},\ \Eprint
  {http://arxiv.org/abs/2009.00026} {arXiv:2009.00026 [hep-ex]} \BibitemShut
  {NoStop}%
\bibitem [{\citenamefont {Zyla}\ \emph {et~al.}(2020)\citenamefont {Zyla} \emph
  {et~al.}}]{ParticleDataGroup:2020ssz}%
  \BibitemOpen
  \bibfield  {author} {\bibinfo {author} {\bibfnamefont {P.~A.}\ \bibnamefont
  {Zyla}} \emph {et~al.} (\bibinfo {collaboration} {Particle Data Group}),\
  }\href {\doibase 10.1093/ptep/ptaa104} {\bibfield  {journal} {\bibinfo
  {journal} {PTEP}\ }\textbf {\bibinfo {volume} {2020}},\ \bibinfo {pages}
  {083C01} (\bibinfo {year} {2020})}\BibitemShut {NoStop}%
\bibitem [{\citenamefont {Duan}\ \emph {et~al.}(2020)\citenamefont {Duan},
  \citenamefont {Luo}, \citenamefont {Liu},\ and\ \citenamefont
  {Matsuki}}]{Duan:2020tsx}%
  \BibitemOpen
  \bibfield  {author} {\bibinfo {author} {\bibfnamefont {M.-X.}\ \bibnamefont
  {Duan}}, \bibinfo {author} {\bibfnamefont {S.-Q.}\ \bibnamefont {Luo}},
  \bibinfo {author} {\bibfnamefont {X.}~\bibnamefont {Liu}}, \ and\ \bibinfo
  {author} {\bibfnamefont {T.}~\bibnamefont {Matsuki}},\ }\href {\doibase
  10.1103/PhysRevD.101.054029} {\bibfield  {journal} {\bibinfo  {journal}
  {Phys. Rev. D}\ }\textbf {\bibinfo {volume} {101}},\ \bibinfo {pages}
  {054029} (\bibinfo {year} {2020})},\ \Eprint
  {http://arxiv.org/abs/2002.03311} {arXiv:2002.03311 [hep-ph]} \BibitemShut
  {NoStop}%
\bibitem [{\citenamefont {Duan}\ \emph {et~al.}(2021)\citenamefont {Duan},
  \citenamefont {Wang}, \citenamefont {Li},\ and\ \citenamefont
  {Liu}}]{Duan:2021bna}%
  \BibitemOpen
  \bibfield  {author} {\bibinfo {author} {\bibfnamefont {M.-X.}\ \bibnamefont
  {Duan}}, \bibinfo {author} {\bibfnamefont {J.-Z.}\ \bibnamefont {Wang}},
  \bibinfo {author} {\bibfnamefont {Y.-S.}\ \bibnamefont {Li}}, \ and\ \bibinfo
  {author} {\bibfnamefont {X.}~\bibnamefont {Liu}},\ }\href {\doibase
  10.1103/PhysRevD.104.034035} {\bibfield  {journal} {\bibinfo  {journal}
  {Phys. Rev. D}\ }\textbf {\bibinfo {volume} {104}},\ \bibinfo {pages}
  {034035} (\bibinfo {year} {2021})},\ \Eprint
  {http://arxiv.org/abs/2104.09132} {arXiv:2104.09132 [hep-ph]} \BibitemShut
  {NoStop}%
\bibitem [{\citenamefont {Li}\ and\ \citenamefont
  {Voloshin}(2015)}]{Li:2015iga}%
  \BibitemOpen
  \bibfield  {author} {\bibinfo {author} {\bibfnamefont {X.}~\bibnamefont
  {Li}}\ and\ \bibinfo {author} {\bibfnamefont {M.~B.}\ \bibnamefont
  {Voloshin}},\ }\href {\doibase 10.1103/PhysRevD.91.114014} {\bibfield
  {journal} {\bibinfo  {journal} {Phys. Rev. D}\ }\textbf {\bibinfo {volume}
  {91}},\ \bibinfo {pages} {114014} (\bibinfo {year} {2015})},\ \Eprint
  {http://arxiv.org/abs/1503.04431} {arXiv:1503.04431 [hep-ph]} \BibitemShut
  {NoStop}%
\bibitem [{\citenamefont {Gonz\'alez}(2017)}]{Gonzalez:2016fsr}%
  \BibitemOpen
  \bibfield  {author} {\bibinfo {author} {\bibfnamefont {P.}~\bibnamefont
  {Gonz\'alez}},\ }\href {\doibase 10.1088/1361-6471/aa6d8a} {\bibfield
  {journal} {\bibinfo  {journal} {J. Phys. G}\ }\textbf {\bibinfo {volume}
  {44}},\ \bibinfo {pages} {075004} (\bibinfo {year} {2017})},\ \Eprint
  {http://arxiv.org/abs/1611.03710} {arXiv:1611.03710 [hep-ph]} \BibitemShut
  {NoStop}%
\bibitem [{\citenamefont {{E. Spadaro Norella and C. Chen
  (LHCb)}}(2022)}]{results}%
  \BibitemOpen
  \bibfield  {author} {\bibinfo {author} {\bibnamefont {{E. Spadaro Norella and
  C. Chen (LHCb)}}},\ }\href@noop {} {\enquote {\bibinfo {title} {Particle zoo
  2.0: New tetra- and pentaquarks at lhcb},}\ }\bibinfo {howpublished}
  {\textsc{url:}~\url{https://indico.cern.ch/event/1176505}} (\bibinfo {year}
  {2022})\BibitemShut {NoStop}%
\bibitem [{\citenamefont {Barnes}\ \emph {et~al.}(2005)\citenamefont {Barnes},
  \citenamefont {Godfrey},\ and\ \citenamefont {Swanson}}]{Barnes:2005pb}%
  \BibitemOpen
  \bibfield  {author} {\bibinfo {author} {\bibfnamefont {T.}~\bibnamefont
  {Barnes}}, \bibinfo {author} {\bibfnamefont {S.}~\bibnamefont {Godfrey}}, \
  and\ \bibinfo {author} {\bibfnamefont {E.~S.}\ \bibnamefont {Swanson}},\
  }\href {\doibase 10.1103/PhysRevD.72.054026} {\bibfield  {journal} {\bibinfo
  {journal} {Phys. Rev. D}\ }\textbf {\bibinfo {volume} {72}},\ \bibinfo
  {pages} {054026} (\bibinfo {year} {2005})},\ \Eprint
  {http://arxiv.org/abs/hep-ph/0505002} {arXiv:hep-ph/0505002} \BibitemShut
  {NoStop}%
\bibitem [{\citenamefont {Li}\ and\ \citenamefont {Chao}(2009)}]{Li:2009zu}%
  \BibitemOpen
  \bibfield  {author} {\bibinfo {author} {\bibfnamefont {B.-Q.}\ \bibnamefont
  {Li}}\ and\ \bibinfo {author} {\bibfnamefont {K.-T.}\ \bibnamefont {Chao}},\
  }\href {\doibase 10.1103/PhysRevD.79.094004} {\bibfield  {journal} {\bibinfo
  {journal} {Phys. Rev. D}\ }\textbf {\bibinfo {volume} {79}},\ \bibinfo
  {pages} {094004} (\bibinfo {year} {2009})},\ \Eprint
  {http://arxiv.org/abs/0903.5506} {arXiv:0903.5506 [hep-ph]} \BibitemShut
  {NoStop}%
\bibitem [{\citenamefont {Gamermann}\ \emph {et~al.}(2007)\citenamefont
  {Gamermann}, \citenamefont {Oset}, \citenamefont {Strottman},\ and\
  \citenamefont {Vicente~Vacas}}]{Gamermann:2006nm}%
  \BibitemOpen
  \bibfield  {author} {\bibinfo {author} {\bibfnamefont {D.}~\bibnamefont
  {Gamermann}}, \bibinfo {author} {\bibfnamefont {E.}~\bibnamefont {Oset}},
  \bibinfo {author} {\bibfnamefont {D.}~\bibnamefont {Strottman}}, \ and\
  \bibinfo {author} {\bibfnamefont {M.~J.}\ \bibnamefont {Vicente~Vacas}},\
  }\href {\doibase 10.1103/PhysRevD.76.074016} {\bibfield  {journal} {\bibinfo
  {journal} {Phys. Rev. D}\ }\textbf {\bibinfo {volume} {76}},\ \bibinfo
  {pages} {074016} (\bibinfo {year} {2007})},\ \Eprint
  {http://arxiv.org/abs/hep-ph/0612179} {arXiv:hep-ph/0612179} \BibitemShut
  {NoStop}%
\bibitem [{\citenamefont {Nieves}\ and\ \citenamefont
  {Valderrama}(2012)}]{Nieves:2012tt}%
  \BibitemOpen
  \bibfield  {author} {\bibinfo {author} {\bibfnamefont {J.}~\bibnamefont
  {Nieves}}\ and\ \bibinfo {author} {\bibfnamefont {M.~P.}\ \bibnamefont
  {Valderrama}},\ }\href {\doibase 10.1103/PhysRevD.86.056004} {\bibfield
  {journal} {\bibinfo  {journal} {Phys. Rev. D}\ }\textbf {\bibinfo {volume}
  {86}},\ \bibinfo {pages} {056004} (\bibinfo {year} {2012})},\ \Eprint
  {http://arxiv.org/abs/1204.2790} {arXiv:1204.2790 [hep-ph]} \BibitemShut
  {NoStop}%
\bibitem [{\citenamefont {Hidalgo-Duque}\ \emph {et~al.}(2013)\citenamefont
  {Hidalgo-Duque}, \citenamefont {Nieves},\ and\ \citenamefont
  {Valderrama}}]{Hidalgo-Duque:2012rqv}%
  \BibitemOpen
  \bibfield  {author} {\bibinfo {author} {\bibfnamefont {C.}~\bibnamefont
  {Hidalgo-Duque}}, \bibinfo {author} {\bibfnamefont {J.}~\bibnamefont
  {Nieves}}, \ and\ \bibinfo {author} {\bibfnamefont {M.~P.}\ \bibnamefont
  {Valderrama}},\ }\href {\doibase 10.1103/PhysRevD.87.076006} {\bibfield
  {journal} {\bibinfo  {journal} {Phys. Rev. D}\ }\textbf {\bibinfo {volume}
  {87}},\ \bibinfo {pages} {076006} (\bibinfo {year} {2013})},\ \Eprint
  {http://arxiv.org/abs/1210.5431} {arXiv:1210.5431 [hep-ph]} \BibitemShut
  {NoStop}%
\bibitem [{\citenamefont {Prelovsek}\ \emph {et~al.}(2021)\citenamefont
  {Prelovsek}, \citenamefont {Collins}, \citenamefont {Mohler}, \citenamefont
  {Padmanath},\ and\ \citenamefont {Piemonte}}]{Prelovsek:2020eiw}%
  \BibitemOpen
  \bibfield  {author} {\bibinfo {author} {\bibfnamefont {S.}~\bibnamefont
  {Prelovsek}}, \bibinfo {author} {\bibfnamefont {S.}~\bibnamefont {Collins}},
  \bibinfo {author} {\bibfnamefont {D.}~\bibnamefont {Mohler}}, \bibinfo
  {author} {\bibfnamefont {M.}~\bibnamefont {Padmanath}}, \ and\ \bibinfo
  {author} {\bibfnamefont {S.}~\bibnamefont {Piemonte}},\ }\href {\doibase
  10.1007/JHEP06(2021)035} {\bibfield  {journal} {\bibinfo  {journal} {JHEP}\
  }\textbf {\bibinfo {volume} {06}},\ \bibinfo {pages} {035} (\bibinfo {year}
  {2021})},\ \Eprint {http://arxiv.org/abs/2011.02542} {arXiv:2011.02542
  [hep-lat]} \BibitemShut {NoStop}%
\bibitem [{\citenamefont {Meng}\ \emph {et~al.}(2021)\citenamefont {Meng},
  \citenamefont {Wang},\ and\ \citenamefont {Zhu}}]{Meng:2020cbk}%
  \BibitemOpen
  \bibfield  {author} {\bibinfo {author} {\bibfnamefont {L.}~\bibnamefont
  {Meng}}, \bibinfo {author} {\bibfnamefont {B.}~\bibnamefont {Wang}}, \ and\
  \bibinfo {author} {\bibfnamefont {S.-L.}\ \bibnamefont {Zhu}},\ }\href
  {\doibase 10.1016/j.scib.2021.03.016} {\bibfield  {journal} {\bibinfo
  {journal} {Sci. Bull.}\ }\textbf {\bibinfo {volume} {66}},\ \bibinfo {pages}
  {1288} (\bibinfo {year} {2021})},\ \Eprint {http://arxiv.org/abs/2012.09813}
  {arXiv:2012.09813 [hep-ph]} \BibitemShut {NoStop}%
\bibitem [{\citenamefont {Dong}\ \emph {et~al.}(2021)\citenamefont {Dong},
  \citenamefont {Guo},\ and\ \citenamefont {Zou}}]{Dong:2021juy}%
  \BibitemOpen
  \bibfield  {author} {\bibinfo {author} {\bibfnamefont {X.-K.}\ \bibnamefont
  {Dong}}, \bibinfo {author} {\bibfnamefont {F.-K.}\ \bibnamefont {Guo}}, \
  and\ \bibinfo {author} {\bibfnamefont {B.-S.}\ \bibnamefont {Zou}},\ }\href
  {\doibase 10.13725/j.cnki.pip.2021.02.001} {\bibfield  {journal} {\bibinfo
  {journal} {Progr. Phys.}\ }\textbf {\bibinfo {volume} {41}},\ \bibinfo
  {pages} {65} (\bibinfo {year} {2021})},\ \Eprint
  {http://arxiv.org/abs/2101.01021} {arXiv:2101.01021 [hep-ph]} \BibitemShut
  {NoStop}%
\bibitem [{\citenamefont {Bayar}\ \emph {et~al.}(2022)\citenamefont {Bayar},
  \citenamefont {Feijoo},\ and\ \citenamefont {Oset}}]{Bayar:2022dqa}%
  \BibitemOpen
  \bibfield  {author} {\bibinfo {author} {\bibfnamefont {M.}~\bibnamefont
  {Bayar}}, \bibinfo {author} {\bibfnamefont {A.}~\bibnamefont {Feijoo}}, \
  and\ \bibinfo {author} {\bibfnamefont {E.}~\bibnamefont {Oset}},\ }\href@noop
  {} {\  (\bibinfo {year} {2022})},\ \Eprint {http://arxiv.org/abs/2207.08490}
  {arXiv:2207.08490 [hep-ph]} \BibitemShut {NoStop}%
\bibitem [{\citenamefont {Ji}\ \emph {et~al.}(2022)\citenamefont {Ji},
  \citenamefont {Dong}, \citenamefont {Albaladejo}, \citenamefont {Du},
  \citenamefont {Guo},\ and\ \citenamefont {Nieves}}]{Ji:2022uie}%
  \BibitemOpen
  \bibfield  {author} {\bibinfo {author} {\bibfnamefont {T.}~\bibnamefont
  {Ji}}, \bibinfo {author} {\bibfnamefont {X.-K.}\ \bibnamefont {Dong}},
  \bibinfo {author} {\bibfnamefont {M.}~\bibnamefont {Albaladejo}}, \bibinfo
  {author} {\bibfnamefont {M.-L.}\ \bibnamefont {Du}}, \bibinfo {author}
  {\bibfnamefont {F.-K.}\ \bibnamefont {Guo}}, \ and\ \bibinfo {author}
  {\bibfnamefont {J.}~\bibnamefont {Nieves}},\ }\href {\doibase
  10.1103/PhysRevD.106.094002} {\bibfield  {journal} {\bibinfo  {journal}
  {Phys. Rev. D}\ }\textbf {\bibinfo {volume} {106}},\ \bibinfo {pages}
  {094002} (\bibinfo {year} {2022})},\ \Eprint
  {http://arxiv.org/abs/2207.08563} {arXiv:2207.08563 [hep-ph]} \BibitemShut
  {NoStop}%
\bibitem [{\citenamefont {Xin}\ \emph {et~al.}(2022)\citenamefont {Xin},
  \citenamefont {Wang},\ and\ \citenamefont {Yang}}]{Xin:2022bzt}%
  \BibitemOpen
  \bibfield  {author} {\bibinfo {author} {\bibfnamefont {Q.}~\bibnamefont
  {Xin}}, \bibinfo {author} {\bibfnamefont {Z.-G.}\ \bibnamefont {Wang}}, \
  and\ \bibinfo {author} {\bibfnamefont {X.-S.}\ \bibnamefont {Yang}},\
  }\href@noop {} {\  (\bibinfo {year} {2022})},\ \Eprint
  {http://arxiv.org/abs/2207.09910} {arXiv:2207.09910 [hep-ph]} \BibitemShut
  {NoStop}%
\bibitem [{\citenamefont {Xie}\ \emph {et~al.}(2022)\citenamefont {Xie},
  \citenamefont {Liu},\ and\ \citenamefont {Geng}}]{Xie:2022lyw}%
  \BibitemOpen
  \bibfield  {author} {\bibinfo {author} {\bibfnamefont {J.-M.}\ \bibnamefont
  {Xie}}, \bibinfo {author} {\bibfnamefont {M.-Z.}\ \bibnamefont {Liu}}, \ and\
  \bibinfo {author} {\bibfnamefont {L.-S.}\ \bibnamefont {Geng}},\ }\href@noop
  {} {\  (\bibinfo {year} {2022})},\ \Eprint {http://arxiv.org/abs/2207.12178}
  {arXiv:2207.12178 [hep-ph]} \BibitemShut {NoStop}%
\bibitem [{\citenamefont {Guo}\ \emph {et~al.}(2022{\natexlab{a}})\citenamefont
  {Guo}, \citenamefont {Li}, \citenamefont {Zhao},\ and\ \citenamefont
  {He}}]{Guo:2022crh}%
  \BibitemOpen
  \bibfield  {author} {\bibinfo {author} {\bibfnamefont {T.}~\bibnamefont
  {Guo}}, \bibinfo {author} {\bibfnamefont {J.}~\bibnamefont {Li}}, \bibinfo
  {author} {\bibfnamefont {J.}~\bibnamefont {Zhao}}, \ and\ \bibinfo {author}
  {\bibfnamefont {L.}~\bibnamefont {He}},\ }\href@noop {} {\  (\bibinfo {year}
  {2022}{\natexlab{a}})},\ \Eprint {http://arxiv.org/abs/2211.10834}
  {arXiv:2211.10834 [hep-ph]} \BibitemShut {NoStop}%
\bibitem [{\citenamefont {Guo}\ \emph {et~al.}(2022{\natexlab{b}})\citenamefont
  {Guo}, \citenamefont {Wang}, \citenamefont {Chen},\ and\ \citenamefont
  {Liu}}]{Guo:2022zbc}%
  \BibitemOpen
  \bibfield  {author} {\bibinfo {author} {\bibfnamefont {D.}~\bibnamefont
  {Guo}}, \bibinfo {author} {\bibfnamefont {J.-Z.}\ \bibnamefont {Wang}},
  \bibinfo {author} {\bibfnamefont {D.-Y.}\ \bibnamefont {Chen}}, \ and\
  \bibinfo {author} {\bibfnamefont {X.}~\bibnamefont {Liu}},\ }\href {\doibase
  10.1103/PhysRevD.106.094037} {\bibfield  {journal} {\bibinfo  {journal}
  {Phys. Rev. D}\ }\textbf {\bibinfo {volume} {106}},\ \bibinfo {pages}
  {094037} (\bibinfo {year} {2022}{\natexlab{b}})},\ \Eprint
  {http://arxiv.org/abs/2210.16720} {arXiv:2210.16720 [hep-ph]} \BibitemShut
  {NoStop}%
\bibitem [{\citenamefont {Chen}\ and\ \citenamefont
  {Huang}(2022)}]{Chen:2022dad}%
  \BibitemOpen
  \bibfield  {author} {\bibinfo {author} {\bibfnamefont {R.}~\bibnamefont
  {Chen}}\ and\ \bibinfo {author} {\bibfnamefont {Q.}~\bibnamefont {Huang}},\
  }\href@noop {} {\  (\bibinfo {year} {2022})},\ \Eprint
  {http://arxiv.org/abs/2209.05180} {arXiv:2209.05180 [hep-ph]} \BibitemShut
  {NoStop}%
\bibitem [{\citenamefont {Shifman}\ \emph
  {et~al.}(1979{\natexlab{a}})\citenamefont {Shifman}, \citenamefont
  {Vainshtein},\ and\ \citenamefont {Zakharov}}]{Shifman:1978bx}%
  \BibitemOpen
  \bibfield  {author} {\bibinfo {author} {\bibfnamefont {M.~A.}\ \bibnamefont
  {Shifman}}, \bibinfo {author} {\bibfnamefont {A.~I.}\ \bibnamefont
  {Vainshtein}}, \ and\ \bibinfo {author} {\bibfnamefont {V.~I.}\ \bibnamefont
  {Zakharov}},\ }\href {\doibase 10.1016/0550-3213(79)90022-1} {\bibfield
  {journal} {\bibinfo  {journal} {Nucl. Phys. B}\ }\textbf {\bibinfo {volume}
  {147}},\ \bibinfo {pages} {385} (\bibinfo {year}
  {1979}{\natexlab{a}})}\BibitemShut {NoStop}%
\bibitem [{\citenamefont {Shifman}\ \emph
  {et~al.}(1979{\natexlab{b}})\citenamefont {Shifman}, \citenamefont
  {Vainshtein},\ and\ \citenamefont {Zakharov}}]{Shifman:1978by}%
  \BibitemOpen
  \bibfield  {author} {\bibinfo {author} {\bibfnamefont {M.~A.}\ \bibnamefont
  {Shifman}}, \bibinfo {author} {\bibfnamefont {A.~I.}\ \bibnamefont
  {Vainshtein}}, \ and\ \bibinfo {author} {\bibfnamefont {V.~I.}\ \bibnamefont
  {Zakharov}},\ }\href {\doibase 10.1016/0550-3213(79)90023-3} {\bibfield
  {journal} {\bibinfo  {journal} {Nucl. Phys. B}\ }\textbf {\bibinfo {volume}
  {147}},\ \bibinfo {pages} {448} (\bibinfo {year}
  {1979}{\natexlab{b}})}\BibitemShut {NoStop}%
\bibitem [{\citenamefont {Lee}\ \emph {et~al.}(2005)\citenamefont {Lee},
  \citenamefont {Kim},\ and\ \citenamefont {Kwon}}]{Lee:2004xk}%
  \BibitemOpen
  \bibfield  {author} {\bibinfo {author} {\bibfnamefont {S.~H.}\ \bibnamefont
  {Lee}}, \bibinfo {author} {\bibfnamefont {H.}~\bibnamefont {Kim}}, \ and\
  \bibinfo {author} {\bibfnamefont {Y.}~\bibnamefont {Kwon}},\ }\href {\doibase
  10.1016/j.physletb.2005.01.029} {\bibfield  {journal} {\bibinfo  {journal}
  {Phys. Lett. B}\ }\textbf {\bibinfo {volume} {609}},\ \bibinfo {pages} {252}
  (\bibinfo {year} {2005})},\ \Eprint {http://arxiv.org/abs/hep-ph/0411104}
  {arXiv:hep-ph/0411104} \BibitemShut {NoStop}%
\bibitem [{\citenamefont {Wang}(2015)}]{Wang:2015nwa}%
  \BibitemOpen
  \bibfield  {author} {\bibinfo {author} {\bibfnamefont {Z.-G.}\ \bibnamefont
  {Wang}},\ }\href {\doibase 10.1142/S0217751X15501687} {\bibfield  {journal}
  {\bibinfo  {journal} {Int. J. Mod. Phys. A}\ }\textbf {\bibinfo {volume}
  {30}},\ \bibinfo {pages} {1550168} (\bibinfo {year} {2015})},\ \Eprint
  {http://arxiv.org/abs/1502.01459} {arXiv:1502.01459 [hep-ph]} \BibitemShut
  {NoStop}%
\bibitem [{\citenamefont {Sundu}\ \emph {et~al.}(2019)\citenamefont {Sundu},
  \citenamefont {Agaev},\ and\ \citenamefont {Azizi}}]{Sundu:2018nxt}%
  \BibitemOpen
  \bibfield  {author} {\bibinfo {author} {\bibfnamefont {H.}~\bibnamefont
  {Sundu}}, \bibinfo {author} {\bibfnamefont {S.~S.}\ \bibnamefont {Agaev}}, \
  and\ \bibinfo {author} {\bibfnamefont {K.}~\bibnamefont {Azizi}},\ }\href
  {\doibase 10.1140/epjc/s10052-019-6737-0} {\bibfield  {journal} {\bibinfo
  {journal} {Eur. Phys. J. C}\ }\textbf {\bibinfo {volume} {79}},\ \bibinfo
  {pages} {215} (\bibinfo {year} {2019})},\ \Eprint
  {http://arxiv.org/abs/1812.10094} {arXiv:1812.10094 [hep-ph]} \BibitemShut
  {NoStop}%
\bibitem [{\citenamefont {Agaev}\ \emph {et~al.}(2019)\citenamefont {Agaev},
  \citenamefont {Azizi}, \citenamefont {Barsbay},\ and\ \citenamefont
  {Sundu}}]{Agaev:2018vag}%
  \BibitemOpen
  \bibfield  {author} {\bibinfo {author} {\bibfnamefont {S.~S.}\ \bibnamefont
  {Agaev}}, \bibinfo {author} {\bibfnamefont {K.}~\bibnamefont {Azizi}},
  \bibinfo {author} {\bibfnamefont {B.}~\bibnamefont {Barsbay}}, \ and\
  \bibinfo {author} {\bibfnamefont {H.}~\bibnamefont {Sundu}},\ }\href
  {\doibase 10.1016/j.nuclphysb.2018.12.021} {\bibfield  {journal} {\bibinfo
  {journal} {Nucl. Phys. B}\ }\textbf {\bibinfo {volume} {939}},\ \bibinfo
  {pages} {130} (\bibinfo {year} {2019})},\ \Eprint
  {http://arxiv.org/abs/1806.04447} {arXiv:1806.04447 [hep-ph]} \BibitemShut
  {NoStop}%
\bibitem [{\citenamefont {Wang}(2020{\natexlab{a}})}]{Wang:2020cme}%
  \BibitemOpen
  \bibfield  {author} {\bibinfo {author} {\bibfnamefont {Z.-G.}\ \bibnamefont
  {Wang}},\ }\href {\doibase 10.1103/PhysRevD.101.074011} {\bibfield  {journal}
  {\bibinfo  {journal} {Phys. Rev. D}\ }\textbf {\bibinfo {volume} {101}},\
  \bibinfo {pages} {074011} (\bibinfo {year} {2020}{\natexlab{a}})},\ \Eprint
  {http://arxiv.org/abs/2001.04095} {arXiv:2001.04095 [hep-ph]} \BibitemShut
  {NoStop}%
\bibitem [{\citenamefont {Wang}(2020{\natexlab{b}})}]{Wang:2020eew}%
  \BibitemOpen
  \bibfield  {author} {\bibinfo {author} {\bibfnamefont {Z.-G.}\ \bibnamefont
  {Wang}},\ }\href@noop {} {\  (\bibinfo {year} {2020}{\natexlab{b}})},\
  \Eprint {http://arxiv.org/abs/2005.12735} {arXiv:2005.12735 [hep-ph]}
  \BibitemShut {NoStop}%
\bibitem [{\citenamefont {Wilson}(1969)}]{Wilson:1969zs}%
  \BibitemOpen
  \bibfield  {author} {\bibinfo {author} {\bibfnamefont {K.~G.}\ \bibnamefont
  {Wilson}},\ }\href {\doibase 10.1103/PhysRev.179.1499} {\bibfield  {journal}
  {\bibinfo  {journal} {Phys. Rev.}\ }\textbf {\bibinfo {volume} {179}},\
  \bibinfo {pages} {1499} (\bibinfo {year} {1969})}\BibitemShut {NoStop}%
\bibitem [{\citenamefont {Workman}\ \emph {et~al.}(2022)\citenamefont {Workman}
  \emph {et~al.}}]{Workman:2022ynf}%
  \BibitemOpen
  \bibfield  {author} {\bibinfo {author} {\bibfnamefont {R.~L.}\ \bibnamefont
  {Workman}} \emph {et~al.} (\bibinfo {collaboration} {Particle Data Group}),\
  }\href {\doibase 10.1093/ptep/ptac097} {\bibfield  {journal} {\bibinfo
  {journal} {PTEP}\ }\textbf {\bibinfo {volume} {2022}},\ \bibinfo {pages}
  {083C01} (\bibinfo {year} {2022})}\BibitemShut {NoStop}%
\bibitem [{\citenamefont {Agaev}\ \emph {et~al.}(2020)\citenamefont {Agaev},
  \citenamefont {Azizi},\ and\ \citenamefont {Sundu}}]{Agaev:2020zad}%
  \BibitemOpen
  \bibfield  {author} {\bibinfo {author} {\bibfnamefont {S.}~\bibnamefont
  {Agaev}}, \bibinfo {author} {\bibfnamefont {K.}~\bibnamefont {Azizi}}, \ and\
  \bibinfo {author} {\bibfnamefont {H.}~\bibnamefont {Sundu}},\ }\href
  {\doibase 10.3906/fiz-2003-15} {\bibfield  {journal} {\bibinfo  {journal}
  {Turk. J. Phys.}\ }\textbf {\bibinfo {volume} {44}},\ \bibinfo {pages} {95}
  (\bibinfo {year} {2020})},\ \Eprint {http://arxiv.org/abs/2004.12079}
  {arXiv:2004.12079 [hep-ph]} \BibitemShut {NoStop}%
\bibitem [{\citenamefont {Wang}\ \emph {et~al.}(2021)\citenamefont {Wang},
  \citenamefont {Li}, \citenamefont {Liang},\ and\ \citenamefont
  {Oset}}]{Wang:2020elp}%
  \BibitemOpen
  \bibfield  {author} {\bibinfo {author} {\bibfnamefont {E.}~\bibnamefont
  {Wang}}, \bibinfo {author} {\bibfnamefont {H.-S.}\ \bibnamefont {Li}},
  \bibinfo {author} {\bibfnamefont {W.-H.}\ \bibnamefont {Liang}}, \ and\
  \bibinfo {author} {\bibfnamefont {E.}~\bibnamefont {Oset}},\ }\href {\doibase
  10.1103/PhysRevD.103.054008} {\bibfield  {journal} {\bibinfo  {journal}
  {Phys. Rev. D}\ }\textbf {\bibinfo {volume} {103}},\ \bibinfo {pages}
  {054008} (\bibinfo {year} {2021})},\ \Eprint
  {http://arxiv.org/abs/2010.15431} {arXiv:2010.15431 [hep-ph]} \BibitemShut
  {NoStop}%
\bibitem [{\citenamefont {Deineka}\ \emph {et~al.}(2022)\citenamefont
  {Deineka}, \citenamefont {Danilkin},\ and\ \citenamefont
  {Vanderhaeghen}}]{Deineka:2021aeu}%
  \BibitemOpen
  \bibfield  {author} {\bibinfo {author} {\bibfnamefont {O.}~\bibnamefont
  {Deineka}}, \bibinfo {author} {\bibfnamefont {I.}~\bibnamefont {Danilkin}}, \
  and\ \bibinfo {author} {\bibfnamefont {M.}~\bibnamefont {Vanderhaeghen}},\
  }\href {\doibase 10.1016/j.physletb.2022.136982} {\bibfield  {journal}
  {\bibinfo  {journal} {Phys. Lett. B}\ }\textbf {\bibinfo {volume} {827}},\
  \bibinfo {pages} {136982} (\bibinfo {year} {2022})},\ \Eprint
  {http://arxiv.org/abs/2111.15033} {arXiv:2111.15033 [hep-ph]} \BibitemShut
  {NoStop}%
\end{thebibliography}%

%\bibliography{ZcbMM}

\end{document}